# Solar Activity Studies using Microwave Imaging Observations*

N. Gopalswamy
Solar Physics Laboratory, Heliophysics Division
NASA Goddard Space Flight Center
Greenbelt, Maryland, USA
nat.gopalswamy@nasa.gov

*Abstract*—We report on the status of solar cycle 24 based on polar prominence eruptions (PEs) and microwave brightness enhancement (MBE) information obtained by the Nobeyama radioheliograph. The north polar region of the Sun had near-zero field strength for more than three years (2012-2015) and ended only in September 2015 as indicated by the presence of polar PEs and the lack of MBE. The zero-polar-field condition in the south started only around 2013, but it ended by June 2014. Thus the asymmetry in the times of polarity reversal switched between cycle 23 and 24. The polar MBE is a good proxy for the polar magnetic field strength as indicated by the high degree of correlation between the two. The cross-correlation between the high- and low-latitude MBEs is significant for a lag of ~5.5 to 7.3 years, suggesting that the polar field of one cycle indicates the sunspot number of the next cycle in agreement with the Babcock-Leighton mechanism of solar cycles. The extended period of near-zero field in the north-polar region should result in a weak and delayed sunspot activity in the northern hemisphere in cycle 25.

## I. Introduction

Interferometric imaging in microwaves helps us understand a number of solar phenomena. The Nobeyama Radioheliograph (NoRH [1]) is one such instrument imaging the Sun at 17 and 34 GHz. The images consist of the following features: (1) the quiet-Sun emission from the upper chromosphere as a uniform disk with a brightness temperature of $10^4$ K. (2) Active regions, which appear as small bright patches due to free-free emission from coronal loops. (3) Sunspot umbrae when the magnetic field is high enough to produce gyroresonance emission. (4) Filaments that appear as dark linear features on the disk because they are cool (~$8 \times 10^3$ K) against the $10^4$ K solar disk. (5) Prominences (regular and eruptive) as bright features above the solar limb against the cold sky in microwaves. (6) Bright active region loops at the limb with a brightness temperature $>10^4$ K, also observed against the cold sky. (7) Bright flare loops are observed when active regions erupt (gyrosynchrotron emission). Weaker post-eruption arcades are observed when quiescent filaments erupt (free-free emission). (8) Coronal dimming during backside eruptions. (9) Weak brightness enhancement in coronal holes.

*Work supported by NASA's Living with a Star TR&T program.

Figure 1 illustrates some of the features noted above. All features on the disk appear bright because their brightness temperatures exceed that of the solar disk. The only exception is the dark filament, which is at a lower temperature than the quiet Sun. Limb features are bright against the optically thin background corona. The locations of prominence eruptions (PEs), the bright active region emission, and the polar microwave brightness enhancement (MBE) are useful for the solar-cycle studies considered in this paper.

When prominences erupt, they can be tracked to 1-2 solar radii (Rs) above the limb within the NoRH field of view [2-3]. The MBE in coronal holes occurs in a narrow wavelength range (0.3 to 3 cm), which includes the 1.7 cm from NoRH corresponding to 17 GHz). MBE is associated with the enhanced unipolar magnetic field inside coronal holes [4-7]. Both PEs and MBE are due to thermal free-free emission from plasmas cooler than the corona (prominences at ~8000 K and the chromosphere at ~10000 K). The low-latitude MBE is thermal free-free emission from the active region corona, with occasional gyro-resonance emission from sunspots [6, 8-10].

In this paper, we obtain the characteristics of solar cycle 24, which happened to be a very small cycle owing to the weak polar field in cycle 23. We compare the high-latitude PE activity from NoRH data with that from the Solar Dynamics Observatory (SDO). We confirm the relation between polar MBE and the polar field strength [6] using the extended NoRH observations. Finally we show that the cross-correlation between the polar low-latitude MBEs is high, supporting the Babcock-Leighton solar cycle mechanism [11-12].

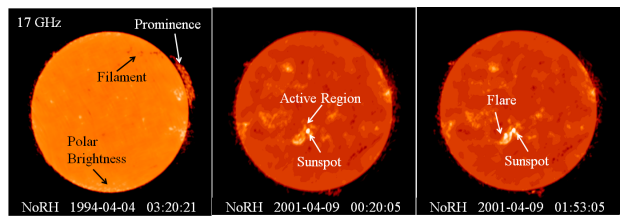

Fig. 1. Three 17 GHz images from NoRH showing dark filaments, bright prominences, active regions, a sunspot in an active region, polar brightness enhancement in coronal hole, and a solar flare. The images contain the solar disk on which various features are superposed. The images have a typical spatial resolution of ~10 arcsec.

## II. HIGH-LATITUDE PROMINENCE ERUPTIONS

High-latitude prominences are known from the 19th century to occur during the maximum phase of the sunspot cycle (see [13] for a review). The polarity reversal at the solar poles was first observed [14] during the maximum of solar cycle 19. The presence of high-latitude filaments was readily confirmed during the cycle-19 maximum phase [15], as was speculated earlier [16]. The locations of PEs detected automatically in the NoRH images were used as a proxy to the high-latitude prominences and filaments to show that the cessation of high-latitude eruptive activity coincided with the establishment of the new-cycle polarity in cycle 23 [17]. Here I confirm the validity of the NoRH results using PEs detected automatically in SDO's Atmospheric Imaging Assembly (AIA, [18]) images at 304 Å.

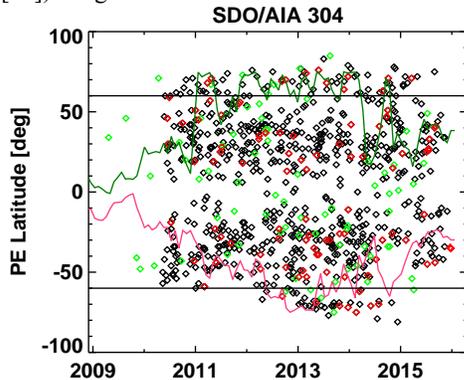

Fig.2. Time-latitude plot of the locations of prominence eruptions (PEs) detected automatically in SDO/AIA images at 304 Å. Black: PEs detected exclusively by SDO, red: PEs detected by both SDO and NoRH, and green: PEs detected only by NoRH. The tilt angle of the heliospheric current sheet is overlaid on the plot (solid curves). The horizontal lines mark ±60º latitudes.

Figure 2 shows the PE locations automatically detected in SDO 304 Å images obtained since 2010 (from SDO launch to March 2016). The SDO 304 Å synoptic images with 2-min cadence were polar-transformed and divided by a background map (pixels with minimum intensity during one day) to get the ratio maps above the limb. The prominence regions were defined as pixels with a ratio >2. Two prominence regions with more than 50% overlapped pixels were considered the same prominence. If the height of a prominence increased monotonically in 5 successive images, it was considered eruptive. All the PEs seen above the limb were detected by the routine, but only PEs with width ≥15º were included in the analysis to eliminate polar jets and other small-scale mass motions [19]. The identifications were also cross-checked with the PEs identified in the NoRH images. The vast majority of PEs in Fig. 2 are from SDO because of the high duty cycle (continuous observations compared to 8-hours of NoRH observations per day). PEs detected by both SDO and NoRH are the second largest in number. There are a significant number of PEs observed exclusively by NoRH probably because of the difference in the appearance in the two wavelengths. EUV 304 Å images are made at a single wavelength, while the microwave PEs are identified in free-free emission, which is thermal emission from the plasma, irrespective of the temperature.

In the northern hemisphere, PEs started occurring at latitudes >60º around the beginning of 2011 and continued for almost four years. In the south, high-latitude PEs started appearing only around the beginning of 2013, indicating a clear asymmetry because sunspot activity started first in the northern hemisphere [6, 17, 20-22]. This evolution of the polar region is mostly captured by the NoRH data except for the last two data points in the two hemispheres and a gap centered around 2012. The time of completion of the polarity reversal, determined by the cessation of high-latitude PEs therefore differ only by a few months. The maximum extent in latitude reached by the computed heliospheric current sheet (HCS) from the Wilcox Solar Observatory (WSO) data has close relationship with the high-latitude PEs. The tilt angle stays above 60º latitudes around the same time as the presence of high-latitude PEs. After the cessation of high-latitude PEs, we can see that the tilt angle dropped to ~30º.

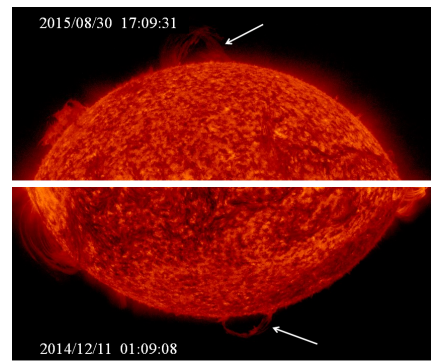

Fig.3. Sections of SDO/AIA images showing the last significant prominence eruptions (pointed by arrows) on 2015 August 30 in the northern hemisphere (top) and on 2014 December 11 in the southern hemisphere (bottom).

The last northern PE event was observed by SDO/AIA between 16:38 and 17:00 UT on 2015 August 30 from N75E05 (see Fig. 3). This PE occurred outside the NoRH observing window. We also confirmed that the PE was not back-sided using STEREO data. The PE was also associated with a narrow CME observed by SOHO (not shown). The last PE from the south was more of a slow rise and disappearance from the backside with no CME. The polar coronal hole was already formed and the PE was located at S81 near the edge of the coronal hole. The PE was not detected in the NoRH images by the automatic detection routine. However, we were able to identify it by examining the NoRH images taken around the time of the eruption. The times of cessation of high-latitude PEs are December 2014 in the south and August 2015 in the north. These times mark the establishment of the new-cycle polarity in the maximum phase of cycle 24. The polarity reversal requires the replacement of the incumbent flux by the unipolar flux of the new polarity. This means all the closed field structures (indicated by the presence of high-latitude prominences) need to be removed by eruptions.

## III. POLAR MICROWAVE BRIGHTNESS ENHANCEMENT

The polar MBE is a direct consequence of the presence of enhanced magnetic field in the chromosphere beneath the polar coronal hole. In addition, the chromospheric layer that is optically thick at 17 GHz must be at a slightly higher temperature, which may also be a consequence of the enhanced polar field B. The microwave butterfly diagram [6] represents the MBE and closely resemble the magnetic butterfly diagram [23]. Figure 4 shows an overlay of the microwave butterfly diagram (contours) on the magnetic butterfly diagram. The high brightness temperature (Tb) contours clearly correspond to the regions of high magnetic field both at high at low latitudes. The extended cycle 23/24 minimum period is evident from the butterfly diagram (2004-2014 in the south and 2002-2012 in the north).

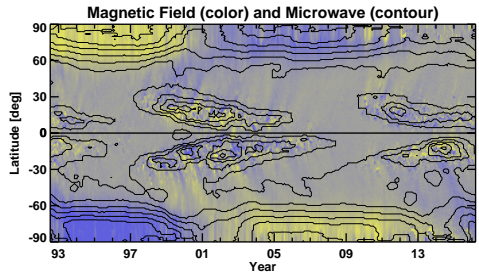

Fig. 4. Microwave butterfly diagram (contours) overlaid on the Kitt Peak magnetic butterfly diagram (June 1992 to February 2016). Yellow and blue colors denote positive and negative polarities, respectively. A 13-rotation smoothing has been used along the time axis to eliminate the periodic variation due to solar B0-angle variation. The contour levels are at 10,000, 10,300, 10,609, 10,927, 11,255, 11,592, and 11,940 K. Flux surges from active regions towards the pole cancel the existing flux before reversal.

The interval of high Tb during the cycle 22/23 minimum is shorter than that during the cycle 23/34 minimum. However, the peak Tb is higher and extends to lower latitudes during the cycle 22/23 minimum. The Tb increase corresponding to the cycle 24/25 minimum has just started in the southern hemisphere, but yet to start in the northern hemisphere. The intervals of Tb close to the quiet Sun values indicate the maximum phases; they vary in length, with the cycle 24 in the north being the longest in recent times [24]. The high-latitude brightness has started increasing by the middle of 2014 in the south, indicating the completion of polarity reversal. In the north, it is not clear, but there is a hint of Tb increase by the end of 2015. In the Tb plots averaged over latitudes >60º, a definite increase in Tb by the end of 2015 was reported [16].

At low latitudes, usually there are multiple intervals of intense activity as indicated by the compact sources in the active region belt. The double peak around the solar maximum is also evident. In cycle 23, activities in both hemispheres contribute to the two peaks. In cycle 24, the first peak is mostly due to the northern hemispheric activity, while the second peak is primarily due to the activity in the southern hemisphere. The weakness of cycle 24 is also clear from the weak Tb, especially in the northern hemisphere.

Figure 5 shows polar B-Tb scatter plots separately for the northern and southern hemispheres. Each data point corresponds to the average value in one Carrington rotation. The polarities are clearly separated in the two hemispheres. The data points in the upper right parts correspond to cycle 22/23 minimum (north: positive; south: negative), when B and Tb were higher. In the cycle 23/24 minimum, the values are lower. This decline in polar activity has been found to be the cause of the weak cycle-24 sunspot activity [25]. The scatter is higher in the northern hemisphere than that in the south. Accordingly, the correlation is higher for the southern hemisphere. The B - Tb relationship in the polar regions reported in [6] was B = 0.0067Tb-72 G, where Tb is in K for the period June 1992 to March 2012. The relation continues to hold for the period extended to the end of 2014 (see the regression equations in Fig. 5).

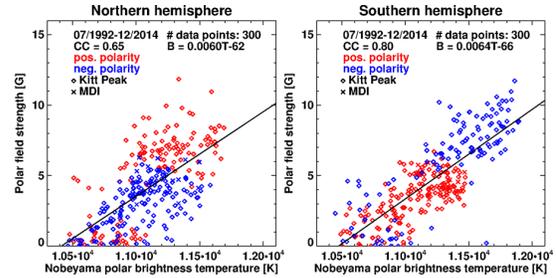

Fig. 5. Scatter plots between the polar magnetic field strength and the microwave brightness temperature averaged over latitudes poleward of 60º in the northern (a) and southern (b) hemispheres. The data are from July 1992 to December 2014. There were two intervals of peak polar B and Tb in each hemisphere as indicated by the colors. Only magnitudes of B are plotted, but the signs are indicated by the colors. Gaps in the Kitt Peak data were filled using data from SOHO's Michelson Doppler Imager (MDI) data (crosses).

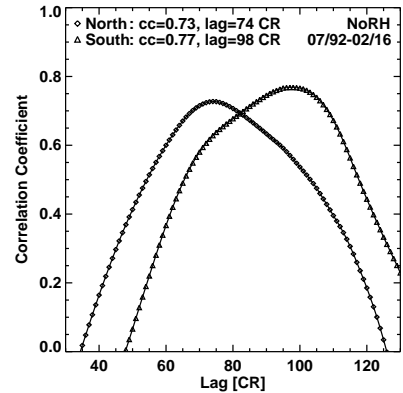

Fig. 6. Cross-correlation between the polar and low-latitude microwave brightness temperatures as a function of time lag between the two data sets in units of Carrington rotations. The correlation coefficients (cc) are shown on the plots.

According to the Babcock-Leighton mechanism, the polar magnetic field in one cycle is advected to the bottom of the convection zone by meridional circulation, which then emerges as the toroidal field at lower latitudes. The polar field strength in one cycle thus indicates the strength of the next activity cycle [25-26]. We performed a cross-correlation analysis between Tb at high and low latitudes as shown in Fig. 6: the correlation is high for a lag of 98 Carrington rotations (~7.3 years) in the north and 74 (5.5 years) in the south. The longer lag in the south is consistent with the longer 23/24 minimum. The correlation coefficients are quite high: 0.73 (north) and 0.77 (south). Thus, the cross-correlation plot is consistent with the Babcock-Leighton mechanism. The onset of cycle 25 is expected to be significantly delayed in the northern hemisphere. On the other hand, cycle 25 is likely to have a normal start in the southern hemisphere but likely to precede the start in the north.

## IV. Summary and Conclusions

I have shown that microwave imaging provides important information on the progression of solar cycles based on the radio emission originating from large-scale features such as eruptive prominences and coronal holes. Microwave emission from active regions is also a good indicator of sunspot activity. The unique appearance of eruptive prominences at high latitudes marks the duration of the maximum phase of the solar cycle. This phase is also indicated by the lack of polar microwave brightness enhancement above the quiet-Sun level. The main conclusions of this study can be summarized as follows.

1. The high-latitude prominence eruptions serve as proxy to the rush-to-the-poles phenomenon. The occurrence of the high-latitude prominence eruptions indicates the presence of bipolar regions and hence is a sign that the polarity reversal has not been completed. 2. The microwave prominence eruptions observed by ground based radio imagers such as the Nobeyama Radioheliograph provide adequate description of the solar activity phases, as confirmed by a comparison with EUV prominence eruptions identified in the Solar Dynamics Observatory images in solar cycle 24. 3. The long-term variability in the polar microwave brightness enhancement is a good indicator of the onset of the maximum phase and the time of polarity reversal. 4. After the reversal, it takes a few more months for the polar coronal holes to fully develop, so some polar prominence eruptions do occur for a few rotations after the reversal. 5. The microwave butterfly diagram indicates the north-south asymmetry in the time of polarity reversal, the extent of the maximum phase, and the strength of active regions. 6. The relation between the polar field strength and the microwave brightness temperature continues to hold when an extended data set is used. Thus, the polar microwave brightness temperature is a good proxy for the polar magnetic field strength. 7. There is a good correlation between the polar microwave brightness in one cycle with the low-latitude brightness in the next, supporting the Babcock-Leighton mechanism. The correlation also shows clear north-south asymmetry. 8. The change in the asymmetry of the polarity reversal in cycle 24 suggests a new asymmetry pattern for solar cycle 25.

## *Acknowledgment*

I thank S. Yashiro and P. Mäkelä for help with some figures. This work benefited from NASA's open data policy in using SDO data. NoRH is currently operated by the Nagoya University in cooperation with the International Consortium for the Continued Operation of the Nobeyama Radioheliograph (ICCON). Magnetic field data were acquired by SOLIS instruments operated by NISP/NSO/AURA/NSF.